\documentclass[a4paper,aps,draft,preprint,nofootinbib,showpacs,showkeys,superscriptaddress,amsfonts,amssymb]{revtex4}

\preprint{SNUTP 02/006}

\begin{document}

\title{~\\\vspace{1cm}\Large\bf Localized nonabelian anomalies at orbifold fixed points}

\author{Florian Gmeiner}
\email[]{gmeiner@th.physik.uni-bonn.de}
\affiliation{Physikalisches Institut, Universit\"at Bonn, Nussallee 12, D-53115 Bonn, Germany}

\author{Jihn E. Kim}
\email[]{jekim@phyp.snu.ac.kr}
\affiliation{School of Physics, Seoul National University, Seoul 151-747, Korea\\\vspace{2cm}}

\author{Hyun Min Lee}
\email[]{minlee@phya.snu.ac.kr}
\affiliation{School of Physics, Seoul National University, Seoul 151-747, Korea\\\vspace{2cm}}

\author{Hans Peter Nilles}
\email[]{nilles@th.physik.uni-bonn.de}
\affiliation{Physikalisches Institut, Universit\"at Bonn, Nussallee 12, D-53115 Bonn, Germany}

\begin{abstract}
Orbifold compactifications of 10D heterotic strings do allow different sets of chiral fermions at
different fixed points. Even if the effective 4D theory is anomaly free by including the bulk
fermions, there arise abelian and nonabelian fixed point anomalies from these chiral fermions.
As the underlying string theory is well defined these localized fixed point anomalies of the chiral
fermions are to be cancelled by a variant of the Green-Schwarz mechanism.\\\vspace{2cm}
\end{abstract}



\maketitle

\newpage

\def\A{{\cal A}}
\def\Q{{\cal Q}}

\newcommand{\debug}{\emph{!!! CHECK !!!}}
\newcommand{\beq}{\begin{equation}}
\newcommand{\eeq}{\end{equation}}
\newcommand{\bea}{\begin{eqnarray}}
\newcommand{\eea}{\end{eqnarray}}
\newcommand{\dd}{\mathrm{d}\,}
\newcommand{\Tr}{\mathrm{Tr}}
\newcommand{\drep}[2]{(\mathbf{#1},\mathbf{#2})}

\section{Introduction}

In the present note we would like to report on the observation of a peculiar type of anomaly
cancellation in the framework of orbifold compactification \cite{Dixon:1985jw,Dixon:1986jc}
of the heterotic $E_8 \times E_8'$ string theory. As an example we consider an orbifold model with
quantized Wilson lines \cite{Ibanez:1987tp} and $SU(3)_c \times SU(3)_w \times U(1)^4 \times SO(12)'
\times U(1)^{'2}$ gauge group, which was previously discussed in the literature \cite{Kim:1993en}.
The anomalies of chiral fermions localized at fixed points (3-branes) do not cancel locally on the
branes, but only globally when the higher dimensions are integrated out. This happens for abelian as
well as nonabelian anomalies. Local anomaly cancellation must be guaranteed by a generalization of the
Green-Schwarz mechanism \cite{Green:1984sg}.

Anomalies represent interesting restrictions in quantum field theory. It is usually assumed that a
consistent theory should not be subject to gauge anomalies. The best argument in this direction is
probably the fact that a family of quarks and leptons in the $SU(3) \times SU(2) \times U(1)$
standard model represents a set of chiral fermions where all gauge anomalies (including mixed gauge and
gravitational anomalies) cancel. In a more general setup (as string theory) the absence of anomalies
is usually obtained as a consequence of quantum consistency (e.g. modular invariance) of the theory.
Nontrivial illustrative examples of such a cancellation are orbifold compactifications where the
chiral fermions in the twisted sectors (required by modular invariance) cancel the anomalies of the
states from the untwisted sector. Moreover, one of the most exciting arguments in favor of string
theory has been an elaborate mechanism of anomaly cancellation discovered by Green and Schwarz
\cite{Green:1984sg}. In $d=10$ it led to the nearly unique choice of gauge group $SO(32)$ or
$E_8 \times E_8$. In the effective low energy $d=10$ field theory this mechanism crucially depends
on the antisymmetric tensor field $B_{MN}$
\footnote{We adopt the following notation: capital indices ($M,N,\ldots$) label all dimensions, greek
indices ($\mu,\nu,\ldots$) the uncompactified ones and lower case latin indices ($m,n,\ldots$) are used
for internal, compactified directions.}
, its field strength
\beq
  H = \dd B+\omega_{YM}-\omega_{L}
\eeq
with Chern-Simons terms
\bea
  \omega_{YM}=\Tr\left(AF-\frac{2}{3}A^3\right), \\\nonumber
  \omega_{L}=\Tr\left(\omega R-\frac{2}{3}\omega^3\right)\nonumber
\eea
and
\beq
  \dd H = \Tr F^2 - \Tr R^2,
\eeq
as well as a specific counterterm
\beq\label{gsterm}
  \int B \wedge I_8,
\eeq
where $I_8$ is a polynomial in $F^2$ and $R^2$ terms \cite{Green:1984sg}. This term gains particular importance in case
of the (apparent) presence of anomalous $U(1)_A$ gauge groups in the effective $d=4$ theory, as the
so-called model independent axion $B_{\mu\nu}$ provides a Higgs mechanism for the ``anomalous'' gauge
boson \cite{Dine:1987bq,Atick:1987gy,Dine:1987gj}.

As string theories are formulated in higher dimensions ($d=10$), there is the possibility that chiral
fermions (and anomalies) might be localized at lower dimensional spaces (branes, fixed points). In
fact, anomalies can appear in branes of even space-time dimension. In the framework of the $d=11$
M-theory picture this is nicely illustrated in the heterotic $E_8 \times E_8$ M-theory of Ho\u{r}ava
and Witten \cite{Horava:1996qa,Horava:1996ma}. The orbifolding procedure of the 11th dimension leads
to (gravitational) anomalies at the $d=10$ boundaries, which require the presence of $E_8$ gauge supermultiplets. The
generalization of the Green-Schwarz mechanism involves in this case the three-index antisymmetric
tensor $C_{MNO}$ and its field strength
\beq
  G = \dd C + \cdots.
\eeq
The generalization of the Green-Schwarz-polynomial (\ref{gsterm}) is partially obtained through the term
\beq
  \int C \wedge G \wedge G
\eeq
in the action of $d=11$ supergravity. In the Ho\u{r}ava-Witten picture this leads to a nontrivial
background (see formula (2.19) of \cite{Horava:1996ma})
\beq\label{GHW}
  G_{MNOP} \sim \epsilon(x^{11})F^2
\eeq
and
\beq
  \dd G_{11MNOP} \sim \delta(x^{11})F^2.
\eeq

We are interested in related questions for a $d=4$ theory embedded in a higher dimensional space-time.
If one starts in a field theoretic example with e.g. $d=5$ space-time, one can obtain chiral fermions
from an orbifolding procedure (as in the case $d=11 \to d=10$ above). Gauge anomalies could appear at
the orbifold fixed points, but not in the 5 dimensional bulk \cite{Arkani-Hamed:2001is}. Such a
picture is common in orbifolds of $d=10$ heterotic $E_8 \times E_8$ string theory as well. As the
simplest example consider the $Z_3$-orbifold of \cite{Dixon:1985jw} with gauge group $SU(3) \times
E_6 \times E_8'$, an untwisted sector with $3\drep{3}{27}$ chiral matter fields and 27 twisted
sectors with matter content $\drep{1}{27}+3\drep{\bar{3}}{1}$ each. The anomalies of the 81 triplets
of $SU(3)$ in the untwisted sector are cancelled by the anti-triplets in the twisted sectors.
Moreover, the contributions to the anomalies are localized at the fixed points and the
$3\drep{\bar{3}}{1}$ at each fixed point cancel the contribution of the untwisted sector
$\frac{1}{N}\cdot 3\drep{3}{27}$, where $N=27$ is the number of fixed points.
The anomalies are thus cancelled locally at each fixed point. In the
present note we want to analyze several questions concerning this local cancellation of anomalies. Is
this the rule or the exception? Is it a property of nonabelian anomalies or does it apply to the
abelian case as well?

We got interested in these questions because of the recent research in higher dimensional (mostly
$d=5$) field theory models of the so-called brane world scenario. Primarily due to
phenomenological motivations one considers matter fields localized at various places, i.e. on branes
or in the higher dimensional bulk. It was observed that in the case of $U(1)$ gauge symmetry,
localized Fayet-Iliopoulos terms and anomalies \cite{Ghilencea:2001bw,Scrucca:2001eb} at $d=4$
fixed points could appear, even if the integrated anomaly vanishes. It was realized that local
counterterms could cancel these localized anomalies \cite{Barbieri:2002ic,Pilo:2002hu,Kim:2002ab}.
Formally this all seems to be consistent (as e.g. explained in \cite{Harvey:2000yg,Callan:1985sa}),
but it remains the question whether this is the correct way to look at the problem, and if yes,
what is the physical origin of the counter terms. Localized anomalies and Fayet-Iliopoulos terms in
fact were found to be at the origin of instabilities \cite{GrootNibbelink:2002wv} of the classical background,
responsible for the localization of some bulk fields at the corresponding branes, rendering the
counter term unphysical through explicit cancellation \cite{Nibbelink:2002qp}.

\section{4D gauge anomaly from 10D Green-Schwarz term}

Answers to these questions could come from embedding the models in string theory with its elaborate cancellation
of anomalies. The Green-Schwarz term in (\ref{gsterm}) plays here the most prominent role. To
illustrate this, let us consider the $E_8 \times E_8'$ heterotic string and the gauge group part of
(\ref{gsterm}), where we consider the first $E_8$ only
\beq\label{gse8}
S_{GS1}=c'\int\, B(\Tr F^2)^2,
\eeq
and where $c'$ is a string theory constant. Then, we can rewrite the
above term after two dimensional integration in the internal
direction $m,n$ of $B_{mn}$ as
\bea
S_{GS1}=2c'\times(aabb)+4c'\times(abab),
\eea
where\footnote{$x=0,\ldots,3$; $y=4,\ldots,9$}
\bea
(aabb)&=&\int d^4 x d^4 y \,a\,\epsilon^{ijkl\mu\nu\rho\sigma}
F^a_{ij}F^a_{kl}F^b_{\mu\nu}F^b_{\rho\sigma} \\
(abab)&=&\int d^4 x d^4 y \,a\,\epsilon^{ijkl\mu\nu\rho\sigma}
F^a_{ij}F^b_{kl}F^a_{\mu\nu}F^b_{\rho\sigma}
\eea
and $a\equiv\int dy^m dy^n B_{mn}$ is a function of the residual
eight coordinates $(x,y)$. Here, since we consider the wrapping
with Wilson lines, $F^a_{ij}$ and $F^a_{kl}$ are assumed to be
functions of $(y^i,y^j)$ and $(y^k,y^l)$, respectively. On the
other hand, $F^b_{\mu\nu}$ is understood to be a function of 4D
space-time. Then, the GS term becomes, after integrating by parts
\bea
S_{GS1}&=&2c'\int d^8 x \,\epsilon^{ijkl\mu\nu\rho\sigma}
\,\bigg[(-\partial_i a A^b_j
+iaf^{bde}A^d_i A^e_j)F^a_{kl}\nonumber \\
&+&(-\partial_k a A^b_l +iaf^{bde}A^d_k A^e_l)F^b_{ij}\bigg]
F^c_{\mu\nu}F^c_{\rho\sigma} \nonumber \\
&+&4c'\times(bcbc)
\eea
where $f^{bde}$ are the group structure constants.
With various background fields for $F$, $A$ and $B$ in extra dimensions this might lead to
counterterms that might be of relevance for anomaly cancellation in the $d=4$ theory, as e.g. in
$d=5$
\beq
  S_{CS}=-\frac{c}{64\pi^2}\int \dd^4 x \dd y \epsilon(y)\epsilon_{MNPQR}A^MF^{NP}F^{PQ}.
\eeq
In the orbifold case the singularity would be of the domain wall type (quite similar to the result
(\ref{GHW}) in the Ho\u{r}ava-Witten-picture).
If we consider (\ref{gse8}) and background fields that differ at various locations in the higher
dimensional theory, one might then also expect various possibilities for anomaly cancellation at
various fixed points.

Another simple example can be found starting from $d=6$ with
\begin{equation}
\int B \wedge F \wedge F.\label{GS6}
\end{equation}
For the $Z_3$ orbifold compactification of a 6D theory, the
relevant singularity is the string type, and the 6D GS-type
interaction (\ref{GS6}) would give a 2D anomaly after the two
dimensional integration, corresponding to the internal directions
$k,l$ of $B_{kl}$. In this case, another 2D integration ($i,j$),
$\int dx^i dx^j$ (i,j=1,2), together with the space-time
integration $\int dx^\mu dx^\nu (\mu,\nu=0,7)$ constitute a 4D
integration, $\int d^4 x$. Indeed, this situation has been
discussed \cite{Callan:1985sa} in 4D with an axion field $a$,
which corresponds to our $\int dy^idy^jB_{ij}$. With the unit
axion decay constant and $U(1)$ field strength $F_{\mu\nu}$, the
action becomes
\begin{equation}
S_{eff}=\frac{e^2}{32\pi^2}\int d^4x
a\epsilon_{ij\mu\nu}F^{ij}F^{\mu\nu} .\label{callan}
\end{equation}
Eq. (\ref{callan}) seems to be gauge invariant, but it is not
with the axionic string background \cite{Callan:1985sa}, and might therefore be relevant for anomaly
cancellation.

Let us now return to the heterotic string in $d=10$ and show, in an instructive example, that such a
mechanism is needed in heterotic string
theory. The model is a $Z_3$ orbifold with two Wilson lines as given in \cite{Kim:1993en}. We shall
try to present here a non-technical description of the model and shall relegate a detailed discussion to
a future publication. We therefore concentrate on the nonabelian gauge groups only and suppress the
(although interesting) discussion of the various $U(1)$ factors.

\section{String orbifold model}

The shift vectors and the two Wilson lines are given by
\begin{eqnarray}
&v= (\frac{1}{3}\ \frac{1}{3}\ \frac{2}{3}\ \frac{1}{3}\
\frac{1}{3}\ \frac{2}{3}\ 0\ \ 0 )\ (0\ \ 0\ \ 0\ \ 0\ \ 0\ \ 0\
\ 0\ \ 0)\nonumber\\
&a_1= (0\ \ 0\ \ 0\ \ \frac{1}{3}\ \frac{1}{3}\ \frac{2}{3}\
\frac{1}{3}\ \frac{1}{3})\
(\frac{1}{3}\ \frac{1}{3}\ \frac{1}{3}\ \frac{1}{3}\ \ 0\ \ 0\ \ 0\ \ 0)\\
&a_3= (0\ \ 0\ \ 0\ \ 0\ \ 0\ \ 0\ \ 0\ \ \frac{2}{3})\
(\frac{1}{3}\ \frac{1}{3}\ \frac{1}{3}\ \frac{1}{3}\ \frac{1}{3}\
\frac{1}{3}\ \frac{1}{3}\ \frac{1}{3}).\nonumber
\end{eqnarray}

The 4D gauge group is $G=SU(3)_c\times SU(3)_w\times
SO(12)^\prime$, where we omitted the $U(1)$ factors. Since
$SO(12)$ does not have anomalies, we focus on the $SU(3)_c\times
SU(3)_w$ anomaly only. The untwisted sector (or bulk) has the
following chiral fermion spectrum with multiplicity 3 due to
$Z_3$,
\begin{equation}
UT: (3^*_c,1),\ \ (1,3^*_w)
\end{equation}
with $c,w$ representing the group. The nine sets of fixed
points\footnote{The multiplicity 3 at each fixed point is
understood.} are labelled by
\begin{eqnarray}
T0&:& v\nonumber\\
T1&:& v+a_1\nonumber\\
T2&:& v-a_1\nonumber\\
T3&:& v+a_3\nonumber\\
T4&:& v-a_3\nonumber\\
T5&:& v+a_1+a_3\nonumber\\
T6&:& v+a_1-a_3\nonumber\\
T7&:& v-a_1+a_3\nonumber\\
T8&:& v-a_1-a_3.\nonumber
\end{eqnarray}
The chiral fermions at each fixed point are
\begin{eqnarray}
T0&:& (3_c,3_w),\ \ 3(3^*_c,1),\ \ 3(1,3^*_w)\nonumber\\
T1&:& (3_c^*,1),\ \ (1,3_w)\nonumber\\
T2&:& (3_c,1),\ \ (1,3^*_w)\nonumber\\
T3&:& (3_c,1),\ \ (3^*_c,1),\ \ (1,3_w),\ \ (1,3^*_w)\nonumber\\
T4&:& (3_c,1),\ \ (3^*_c,1),\ \ (1,3_w),\ \ (1,3^*_w)\nonumber\\
T5&:& (3_c,1),\ \ (3^*_c,1),\ \ (1,3_w),\ \ (1,3^*_w)\nonumber\\
T6&:& (12)^\prime\nonumber\\
T7&:& (3_c,1), \ \ (1,3_w)\nonumber\\
T8&:& (3_c,1),\ \ (3^*_c,1),\ \ (1,3_w),\ \ (1,3^*_w),\nonumber
\end{eqnarray}
where we omitted singlets. Since the representation is exactly
symmetric under $c\leftrightarrow w$, we consider the anomaly
only for one $SU(3)$, for example $SU(3)_c$. Note that fermions
from $T3,T4,T5,T6$ and $T8$, respectively, do not give nonabelian
gauge anomalies. At these fifteen fixed points the bulk fermion
anomaly for $SU(3)_c$ is
\begin{equation}
A_{bulk}=-\frac{3}{27}\Q,
\end{equation}
where we considered the multiplicity 3 and the equal distribution
of the bulk fermion anomaly between the 27 fixed points.
With $\Q$ we denote the standard anomaly contribution from an $SU(3)$ triplet
\beq\
  \Q = -\frac{1}{32\pi^2}\Tr(\{t_a,t_b\}t)\epsilon^{\mu\nu\rho\sigma}F^a_{\mu\nu}F^b_{\rho\sigma}.
\eeq
The 9 fixed points, shown as $Ti (i=0,1,\ldots,8)$, have the multiplicity 3. The
fixed point fermion contributions are $-\Q,\Q$ and $\Q$ at each fixed point of class
$T1,T2$ and $T7$, respectively, and zero at the other fixed
points. Thus, the anomalies localized at the fixed points are
\begin{eqnarray}
T0&:& -\frac{1}{9}\Q,\ \ T1: -\frac{10}{9}\Q,\ \ T2: +\frac{8}{9}\Q\nonumber\\
T3&:& -\frac{1}{9}\Q,\ \ T4: -\frac{1}{9}\Q,\ \ T5: -\frac{1}{9}\Q\label{GSKK}\\
T6&:& -\frac{1}{9}\Q,\ \ T7: +\frac{8}{9}\Q,\ \ T8: -\frac{1}{9}\Q.\nonumber
\end{eqnarray}

We see here explicitly that these localized anomalies need a nontrivial cancellation through a
generalization of a Green-Schwarz \cite{Green:1984sg} or inflow \cite{Callan:1985sa} mechanism.
A detailed discussion is beyond the scope of this paper and will be relegated to a future
publication, where we also present additional explicit examples.

\section{Conclusions}
It remains to be seen whether this appearence of localized fermions has implications for models of
particle physics, that try to generalize the standard model of strong and electroweak interactions.
One might hope, that these nontrivial examples of orbifolds of the heterotic string might be of help
understanding (orbifold) compactifications of the $d=11$ theory of Ho\u{r}ava and Witten. Some
attempts in this direction have been made \cite{Kaplunovsky:1999ia,Faux:1999hm}, for a detailed
discussion see \cite{Gorbatov:2001pw}. We might also hope to see a connection to compactifications
of Type I string theory or Type II orientifolds. Recently there has been a discussion of localized
$U(1)$ anomalies in this framework \cite{Scrucca:2002is,Antoniadis:2002cs}.

A possible relation to the mechanism of anomaly cancellation in M-theory compactifications on
manifolds of $G2$ holonomy \cite{Witten:2001uq} is less apparent, because there the singularities
have to be more severe than those of the orbifold fixed points considered here.

\acknowledgments
Work supported in part by the European Community's Human Potential Programme under contracts
HPRN-CT-2000-00131 Quantum Spacetime, HPRN-CT-2000-00148 Physics Across the Present Energy Frontier
and HPRN-CT-2000-00152 Supersymmetry and the Early Universe. JEK is supported in part by the BK21
program of Ministry of Education, the KOSEF Sundo Grant, and by the Office of
Research Affairs of Seoul National University.

\bibliography{anomaly}

\begin{thebibliography}{26}
\expandafter\ifx\csname natexlab\endcsname\relax\def\natexlab#1{#1}\fi
\expandafter\ifx\csname bibnamefont\endcsname\relax
  \def\bibnamefont#1{#1}\fi
\expandafter\ifx\csname bibfnamefont\endcsname\relax
  \def\bibfnamefont#1{#1}\fi
\expandafter\ifx\csname citenamefont\endcsname\relax
  \def\citenamefont#1{#1}\fi
\expandafter\ifx\csname url\endcsname\relax
  \def\url#1{\texttt{#1}}\fi
\expandafter\ifx\csname urlprefix\endcsname\relax\def\urlprefix{URL }\fi
\providecommand{\bibinfo}[2]{#2}
\providecommand{\eprint}[2][]{\url{#2}}

\bibitem[{\citenamefont{Dixon et~al.}(1985)\citenamefont{Dixon, Harvey, Vafa,
  and Witten}}]{Dixon:1985jw}
\bibinfo{author}{\bibfnamefont{L.~J.} \bibnamefont{Dixon}},
  \bibinfo{author}{\bibfnamefont{J.~A.} \bibnamefont{Harvey}},
  \bibinfo{author}{\bibfnamefont{C.}~\bibnamefont{Vafa}}, \bibnamefont{and}
  \bibinfo{author}{\bibfnamefont{E.}~\bibnamefont{Witten}},
  \bibinfo{journal}{Nucl. Phys.} \textbf{\bibinfo{volume}{B261}},
  \bibinfo{pages}{678} (\bibinfo{year}{1985}).

\bibitem[{\citenamefont{Dixon et~al.}(1986)\citenamefont{Dixon, Harvey, Vafa,
  and Witten}}]{Dixon:1986jc}
\bibinfo{author}{\bibfnamefont{L.~J.} \bibnamefont{Dixon}},
  \bibinfo{author}{\bibfnamefont{J.~A.} \bibnamefont{Harvey}},
  \bibinfo{author}{\bibfnamefont{C.}~\bibnamefont{Vafa}}, \bibnamefont{and}
  \bibinfo{author}{\bibfnamefont{E.}~\bibnamefont{Witten}},
  \bibinfo{journal}{Nucl. Phys.} \textbf{\bibinfo{volume}{B274}},
  \bibinfo{pages}{285} (\bibinfo{year}{1986}).

\bibitem[{\citenamefont{Ibanez et~al.}(1987)\citenamefont{Ibanez, Nilles, and
  Quevedo}}]{Ibanez:1987tp}
\bibinfo{author}{\bibfnamefont{L.~E.} \bibnamefont{Ibanez}},
  \bibinfo{author}{\bibfnamefont{H.~P.} \bibnamefont{Nilles}},
  \bibnamefont{and} \bibinfo{author}{\bibfnamefont{F.}~\bibnamefont{Quevedo}},
  \bibinfo{journal}{Phys. Lett.} \textbf{\bibinfo{volume}{B187}},
  \bibinfo{pages}{25} (\bibinfo{year}{1987}).

\bibitem[{\citenamefont{Kim and Kim}(1993)}]{Kim:1993en}
\bibinfo{author}{\bibfnamefont{H.~B.} \bibnamefont{Kim}} \bibnamefont{and}
  \bibinfo{author}{\bibfnamefont{J.~E.} \bibnamefont{Kim}},
  \bibinfo{journal}{Phys. Lett.} \textbf{\bibinfo{volume}{B300}},
  \bibinfo{pages}{343} (\bibinfo{year}{1993}),
  \eprint[http://arXiv.org/abs]{hep-ph/9212311}.

\bibitem[{\citenamefont{Green and Schwarz}(1984)}]{Green:1984sg}
\bibinfo{author}{\bibfnamefont{M.~B.} \bibnamefont{Green}} \bibnamefont{and}
  \bibinfo{author}{\bibfnamefont{J.~H.} \bibnamefont{Schwarz}},
  \bibinfo{journal}{Phys. Lett.} \textbf{\bibinfo{volume}{B149}},
  \bibinfo{pages}{117} (\bibinfo{year}{1984}).

\bibitem[{\citenamefont{Dine et~al.}(1987{\natexlab{a}})\citenamefont{Dine,
  Seiberg, Wen, and Witten}}]{Dine:1987bq}
\bibinfo{author}{\bibfnamefont{M.}~\bibnamefont{Dine}},
  \bibinfo{author}{\bibfnamefont{N.}~\bibnamefont{Seiberg}},
  \bibinfo{author}{\bibfnamefont{X.~G.} \bibnamefont{Wen}}, \bibnamefont{and}
  \bibinfo{author}{\bibfnamefont{E.}~\bibnamefont{Witten}},
  \bibinfo{journal}{Nucl. Phys.} \textbf{\bibinfo{volume}{B289}},
  \bibinfo{pages}{319} (\bibinfo{year}{1987}{\natexlab{a}}).

\bibitem[{\citenamefont{Atick et~al.}(1987)\citenamefont{Atick, Dixon, and
  Sen}}]{Atick:1987gy}
\bibinfo{author}{\bibfnamefont{J.~J.} \bibnamefont{Atick}},
  \bibinfo{author}{\bibfnamefont{L.~J.} \bibnamefont{Dixon}}, \bibnamefont{and}
  \bibinfo{author}{\bibfnamefont{A.}~\bibnamefont{Sen}},
  \bibinfo{journal}{Nucl. Phys.} \textbf{\bibinfo{volume}{B292}},
  \bibinfo{pages}{109} (\bibinfo{year}{1987}).

\bibitem[{\citenamefont{Dine et~al.}(1987{\natexlab{b}})\citenamefont{Dine,
  Ichinose, and Seiberg}}]{Dine:1987gj}
\bibinfo{author}{\bibfnamefont{M.}~\bibnamefont{Dine}},
  \bibinfo{author}{\bibfnamefont{I.}~\bibnamefont{Ichinose}}, \bibnamefont{and}
  \bibinfo{author}{\bibfnamefont{N.}~\bibnamefont{Seiberg}},
  \bibinfo{journal}{Nucl. Phys.} \textbf{\bibinfo{volume}{B293}},
  \bibinfo{pages}{253} (\bibinfo{year}{1987}{\natexlab{b}}).

\bibitem[{\citenamefont{Horava and Witten}(1996{\natexlab{a}})}]{Horava:1996qa}
\bibinfo{author}{\bibfnamefont{P.}~\bibnamefont{Horava}} \bibnamefont{and}
  \bibinfo{author}{\bibfnamefont{E.}~\bibnamefont{Witten}},
  \bibinfo{journal}{Nucl. Phys.} \textbf{\bibinfo{volume}{B460}},
  \bibinfo{pages}{506} (\bibinfo{year}{1996}{\natexlab{a}}),
  \eprint[http://arXiv.org/abs]{hep-th/9510209}.

\bibitem[{\citenamefont{Horava and Witten}(1996{\natexlab{b}})}]{Horava:1996ma}
\bibinfo{author}{\bibfnamefont{P.}~\bibnamefont{Horava}} \bibnamefont{and}
  \bibinfo{author}{\bibfnamefont{E.}~\bibnamefont{Witten}},
  \bibinfo{journal}{Nucl. Phys.} \textbf{\bibinfo{volume}{B475}},
  \bibinfo{pages}{94} (\bibinfo{year}{1996}{\natexlab{b}}),
  \eprint[http://arXiv.org/abs]{hep-th/9603142}.

\bibitem[{\citenamefont{Arkani-Hamed et~al.}(2001)\citenamefont{Arkani-Hamed,
  Cohen, and Georgi}}]{Arkani-Hamed:2001is}
\bibinfo{author}{\bibfnamefont{N.}~\bibnamefont{Arkani-Hamed}},
  \bibinfo{author}{\bibfnamefont{A.~G.} \bibnamefont{Cohen}}, \bibnamefont{and}
  \bibinfo{author}{\bibfnamefont{H.}~\bibnamefont{Georgi}},
  \bibinfo{journal}{Phys. Lett.} \textbf{\bibinfo{volume}{B516}},
  \bibinfo{pages}{395} (\bibinfo{year}{2001}),
  \eprint[http://arXiv.org/abs]{hep-th/0103135}.

\bibitem[{\citenamefont{Ghilencea et~al.}(2001)\citenamefont{Ghilencea,
  Groot~Nibbelink, and Nilles}}]{Ghilencea:2001bw}
\bibinfo{author}{\bibfnamefont{D.~M.} \bibnamefont{Ghilencea}},
  \bibinfo{author}{\bibfnamefont{S.}~\bibnamefont{Groot~Nibbelink}},
  \bibnamefont{and} \bibinfo{author}{\bibfnamefont{H.~P.}
  \bibnamefont{Nilles}}, \bibinfo{journal}{Nucl. Phys.}
  \textbf{\bibinfo{volume}{B619}}, \bibinfo{pages}{385} (\bibinfo{year}{2001}),
  \eprint[http://arXiv.org/abs]{hep-th/0108184}.

\bibitem[{\citenamefont{Scrucca
  et~al.}(2002{\natexlab{a}})\citenamefont{Scrucca, Serone, Silvestrini, and
  Zwirner}}]{Scrucca:2001eb}
\bibinfo{author}{\bibfnamefont{C.~A.} \bibnamefont{Scrucca}},
  \bibinfo{author}{\bibfnamefont{M.}~\bibnamefont{Serone}},
  \bibinfo{author}{\bibfnamefont{L.}~\bibnamefont{Silvestrini}},
  \bibnamefont{and} \bibinfo{author}{\bibfnamefont{F.}~\bibnamefont{Zwirner}},
  \bibinfo{journal}{Phys. Lett.} \textbf{\bibinfo{volume}{B525}},
  \bibinfo{pages}{169} (\bibinfo{year}{2002}{\natexlab{a}}),
  \eprint[http://arXiv.org/abs]{hep-th/0110073}.

\bibitem[{\citenamefont{Barbieri et~al.}(2002)\citenamefont{Barbieri, Contino,
  Creminelli, Rattazzi, and Scrucca}}]{Barbieri:2002ic}
\bibinfo{author}{\bibfnamefont{R.}~\bibnamefont{Barbieri}},
  \bibinfo{author}{\bibfnamefont{R.}~\bibnamefont{Contino}},
  \bibinfo{author}{\bibfnamefont{P.}~\bibnamefont{Creminelli}},
  \bibinfo{author}{\bibfnamefont{R.}~\bibnamefont{Rattazzi}}, \bibnamefont{and}
  \bibinfo{author}{\bibfnamefont{C.~A.} \bibnamefont{Scrucca}}
  (\bibinfo{year}{2002}), \eprint[http://arXiv.org/abs]{hep-th/0203039}.

\bibitem[{\citenamefont{Pilo and Riotto}(2002)}]{Pilo:2002hu}
\bibinfo{author}{\bibfnamefont{L.}~\bibnamefont{Pilo}} \bibnamefont{and}
  \bibinfo{author}{\bibfnamefont{A.}~\bibnamefont{Riotto}}
  (\bibinfo{year}{2002}), \eprint[http://arXiv.org/abs]{hep-th/0202144}.

\bibitem[{\citenamefont{Kim et~al.}(2002)\citenamefont{Kim, Kim, and
  Lee}}]{Kim:2002ab}
\bibinfo{author}{\bibfnamefont{H.-D.} \bibnamefont{Kim}},
  \bibinfo{author}{\bibfnamefont{J.~E.} \bibnamefont{Kim}}, \bibnamefont{and}
  \bibinfo{author}{\bibfnamefont{H.~M.} \bibnamefont{Lee}}
  (\bibinfo{year}{2002}), \eprint[http://arXiv.org/abs]{hep-th/0204132}.

\bibitem[{\citenamefont{Harvey and Ruchayskiy}(2001)}]{Harvey:2000yg}
\bibinfo{author}{\bibfnamefont{J.~A.} \bibnamefont{Harvey}} \bibnamefont{and}
  \bibinfo{author}{\bibfnamefont{O.}~\bibnamefont{Ruchayskiy}},
  \bibinfo{journal}{JHEP} \textbf{\bibinfo{volume}{06}}, \bibinfo{pages}{044}
  (\bibinfo{year}{2001}), \eprint[http://arXiv.org/abs]{hep-th/0007037}.

\bibitem[{\citenamefont{Callan and Harvey}(1985)}]{Callan:1985sa}
\bibinfo{author}{\bibfnamefont{J.}~\bibnamefont{Callan},
  \bibfnamefont{Curtis~G.}} \bibnamefont{and}
  \bibinfo{author}{\bibfnamefont{J.~A.} \bibnamefont{Harvey}},
  \bibinfo{journal}{Nucl. Phys.} \textbf{\bibinfo{volume}{B250}},
  \bibinfo{pages}{427} (\bibinfo{year}{1985}).

\bibitem[{\citenamefont{Groot~Nibbelink
  et~al.}(2002)\citenamefont{Groot~Nibbelink, Nilles, and
  Olechowski}}]{GrootNibbelink:2002wv}
\bibinfo{author}{\bibfnamefont{S.}~\bibnamefont{Groot~Nibbelink}},
  \bibinfo{author}{\bibfnamefont{H.~P.} \bibnamefont{Nilles}},
  \bibnamefont{and}
  \bibinfo{author}{\bibfnamefont{M.}~\bibnamefont{Olechowski}}
  (\bibinfo{year}{2002}), \eprint[http://arXiv.org/abs]{hep-th/0203055}.

\bibitem[{\citenamefont{Nibbelink et~al.}(2002)\citenamefont{Nibbelink, Nilles,
  and Olechowski}}]{Nibbelink:2002qp}
\bibinfo{author}{\bibfnamefont{S.~G.} \bibnamefont{Nibbelink}},
  \bibinfo{author}{\bibfnamefont{H.~P.} \bibnamefont{Nilles}},
  \bibnamefont{and}
  \bibinfo{author}{\bibfnamefont{M.}~\bibnamefont{Olechowski}}
  (\bibinfo{year}{2002}), \eprint[http://arXiv.org/abs]{hep-th/0205012}.

\bibitem[{\citenamefont{Kaplunovsky et~al.}(2000)\citenamefont{Kaplunovsky,
  Sonnenschein, Theisen, and Yankielowicz}}]{Kaplunovsky:1999ia}
\bibinfo{author}{\bibfnamefont{V.}~\bibnamefont{Kaplunovsky}},
  \bibinfo{author}{\bibfnamefont{J.}~\bibnamefont{Sonnenschein}},
  \bibinfo{author}{\bibfnamefont{S.}~\bibnamefont{Theisen}}, \bibnamefont{and}
  \bibinfo{author}{\bibfnamefont{S.}~\bibnamefont{Yankielowicz}},
  \bibinfo{journal}{Nucl. Phys.} \textbf{\bibinfo{volume}{B590}},
  \bibinfo{pages}{123} (\bibinfo{year}{2000}),
  \eprint[http://arXiv.org/abs]{hep-th/9912144}.

\bibitem[{\citenamefont{Faux et~al.}(1999)\citenamefont{Faux, Lust, and
  Ovrut}}]{Faux:1999hm}
\bibinfo{author}{\bibfnamefont{M.}~\bibnamefont{Faux}},
  \bibinfo{author}{\bibfnamefont{D.}~\bibnamefont{Lust}}, \bibnamefont{and}
  \bibinfo{author}{\bibfnamefont{B.~A.} \bibnamefont{Ovrut}},
  \bibinfo{journal}{Nucl. Phys.} \textbf{\bibinfo{volume}{B554}},
  \bibinfo{pages}{437} (\bibinfo{year}{1999}),
  \eprint[http://arXiv.org/abs]{hep-th/9903028}.

\bibitem[{\citenamefont{Gorbatov et~al.}(2001)\citenamefont{Gorbatov,
  Kaplunovsky, Sonnenschein, Theisen, and Yankielowicz}}]{Gorbatov:2001pw}
\bibinfo{author}{\bibfnamefont{E.}~\bibnamefont{Gorbatov}},
  \bibinfo{author}{\bibfnamefont{V.~S.} \bibnamefont{Kaplunovsky}},
  \bibinfo{author}{\bibfnamefont{J.}~\bibnamefont{Sonnenschein}},
  \bibinfo{author}{\bibfnamefont{S.}~\bibnamefont{Theisen}}, \bibnamefont{and}
  \bibinfo{author}{\bibfnamefont{S.}~\bibnamefont{Yankielowicz}}
  (\bibinfo{year}{2001}), \eprint[http://arXiv.org/abs]{hep-th/0108135}.

\bibitem[{\citenamefont{Scrucca
  et~al.}(2002{\natexlab{b}})\citenamefont{Scrucca, Serone, and
  Trapletti}}]{Scrucca:2002is}
\bibinfo{author}{\bibfnamefont{C.~A.} \bibnamefont{Scrucca}},
  \bibinfo{author}{\bibfnamefont{M.}~\bibnamefont{Serone}}, \bibnamefont{and}
  \bibinfo{author}{\bibfnamefont{M.}~\bibnamefont{Trapletti}}
  (\bibinfo{year}{2002}{\natexlab{b}}),
  \eprint[http://arXiv.org/abs]{hep-th/0203190}.

\bibitem[{\citenamefont{Antoniadis et~al.}(2002)\citenamefont{Antoniadis,
  Kiritsis, and Rizos}}]{Antoniadis:2002cs}
\bibinfo{author}{\bibfnamefont{I.}~\bibnamefont{Antoniadis}},
  \bibinfo{author}{\bibfnamefont{E.}~\bibnamefont{Kiritsis}}, \bibnamefont{and}
  \bibinfo{author}{\bibfnamefont{J.}~\bibnamefont{Rizos}}
  (\bibinfo{year}{2002}), \eprint[http://arXiv.org/abs]{hep-th/0204153}.

\bibitem[{\citenamefont{Witten}(2001)}]{Witten:2001uq}
\bibinfo{author}{\bibfnamefont{E.}~\bibnamefont{Witten}}
  (\bibinfo{year}{2001}), \eprint[http://arXiv.org/abs]{hep-th/0108165}.

\end{thebibliography}

\end{document}